%% file: submitted_6pages.tex
\documentclass[10pt,conference]{IEEEtran}

\usepackage{comment}
\usepackage{booktabs} 

\usepackage{layout}

\usepackage[hyphens]{url}  
\usepackage[hidelinks,bookmarks=false]{hyperref}  

\usepackage{lmodern}  
\usepackage{mathptmx} 

\usepackage[T1]{fontenc}
\usepackage{mathtools}
\usepackage[inline]{enumitem}

\usepackage{mathtools}

\usepackage[nolist]{acronym}

\usepackage{graphicx}
\usepackage[tight,footnotesize]{subfigure}
\usepackage{epstopdf}
\graphicspath{{Figures/}}

\usepackage{subcaption}  

\usepackage{xcolor}
\usepackage{gensymb}

\usepackage{makecell} 

\usepackage{url}
\usepackage{multirow}
\usepackage{balance}
\usepackage[numbers,sort&compress]{natbib}
\usepackage{amsmath}
\usepackage{amssymb}
\usepackage{bm}
\usepackage{lipsum}
\usepackage{color}
\usepackage[table,xcdraw]{xcolor}
\usepackage{threeparttable}

\usepackage{physics} 

\usepackage{multirow}
\usepackage{authblk}

\usepackage[linesnumbered,ruled,vlined]{algorithm2e}

\SetCommentSty{mycommfont}

\SetKwInput{KwInput}{Input}                
\SetKwInput{KwOutput}{Output}              

\renewcommand{\baselinestretch}{.940}

\usepackage{lscape}
\usepackage{longtable}

\begin{document}
\bstctlcite{IEEEexample:BSTcontrol}


\title{Maximising Energy Efficiency in Large-Scale Open RAN: Hybrid xApps and Digital Twin Integration}

\author{
Ahmed Al-Tahmeesschi\textsuperscript{1,*}, Yi Chu\textsuperscript{1,*}, Gurdeep Singh\textsuperscript{2}, Charles Turyagyenda\textsuperscript{2},\\ Dritan Kaleshi\textsuperscript{2},
David Grace\textsuperscript{1}, Hamed Ahmadi\textsuperscript{1}
\vspace{1mm}\\
\textsuperscript{1}{School of Physics, Engineering and Technology, University of York, United Kingdom}\\
\textsuperscript{2}{Digital Catapult, United Kingdom}
}

\maketitle
\IEEEpubidadjcol

\def\thefootnote{*}\footnotetext{These authors contributed equally to this work.}

\begin{abstract}

The growing demand for high-speed, ultra-reliable, and low-latency communications in 5G and beyond networks has significantly driven up power consumption, particularly within the Radio Access Network (RAN). This surge in energy demand poses critical operational and sustainability challenges for mobile network operators, necessitating innovative solutions that enhance energy efficiency without compromising Quality of Service (QoS). Open Radio Access Network (O-RAN), spearheaded by the O-RAN Alliance, offers disaggregated, programmable, and intelligent architectures, promoting flexibility, interoperability, and cost-effectiveness. However, this disaggregated approach adds complexity, particularly in managing power consumption across diverse network components such as Open Radio Units (RUs). In this paper, we propose a hybrid xApp leveraging heuristic methods and unsupervised machine learning, integrated with digital twin technology through the TeraVM AI RAN Scenario Generator (AI-RSG). This approach dynamically manages RU sleep modes to effectively reduce energy consumption. Our experimental evaluation in a realistic, large-scale emulated Open RAN scenario demonstrates that the hybrid xApp achieves approximately 13$\%$ energy savings, highlighting its practicality and significant potential for real-world deployments without compromising user QoS.


\end{abstract}

\IEEEpeerreviewmaketitle


\section{Introduction}
\label{sec:Introduction}
The rapid evolution of wireless communication technologies necessitates innovative approaches to meet the increasing demands for throughput, coverage, and user experiences. However, there are critical environmental impacts arise together with the increasing demands such as energy consumption and carbon footprint. Throughout the generations of the mobile networks, the \ac{RAN} has always been an imperative component and the direct gateway of connecting the mobile \ac{UE} over the air. The RAN includes the computing infrastructure hosting the heavy baseband signal processing as well as power-hungry hardware components (such as the power amplifiers), therefore, its energy consumption has always been a major concern. From a decade ago the \acp{MNO} have already been reported as one of the top energy consumers \cite{feng2012survey}. The concern has not been resolved with the deployment of the \ac{5G}\cite{Imran_EE_survey_2023}. To address the increasing demands for network capacity, coverage and latency, mass deployment of ultra-dense small-scale \acp{BS}, known as network densification, is the major trend of 5G and future networks. Such \ac{HDD} inevitability increases mobile network energy consumption, leading to a greater carbon footprint and higher operating cost for the \acp{MNO}.


The most well-established architecture of Open RAN is \ac{O-RAN} which was proposed by the O-RAN Alliance \cite{bonati2021intelligence}. The benefits of \ac{O-RAN} include standardized open interfaces that allow multi-vendor network deployment \cite{bonati2021intelligence} and the integrated \ac{AI}/\ac{ML} hosted in the \ac{RIC} \cite{ORAN_Architecture_2022}. O-RAN disaggregates the RAN into a \ac{RU}, a \ac{DU} and a \ac{CU} with each unit hosting a certain set of \ac{RAN} functions according to different functional split options \cite{RCRWireless_FunctionalSplits_2021}. The well-defined interfaces allow interoperability across \ac{RU}, \ac{DU} and \ac{CU} from multiple vendors which lowers the difficulty for small and medium manufacturers to enter the \ac{RAN} market, therefore, fostering the market competitiveness, innovation and upgrade cycles \cite{polese2023understanding}. The \ac{RIC} on the other hand, acts as the central intelligence of the \ac{RAN} which conducts various \ac{AI}/\ac{ML} based network performance optimizations via specific interfaces such as E2 \cite{RIC} and O1 \cite{O1}. 

The widely supported E2 interface specifies the messages \cite{E2AP} between the \ac{RAN} and the \ac{Near-RT} \ac{RIC} which handles operations requiring a latency of between 0.1 and 1 second \cite{RIC}. The customisable xApps hosted at the \ac{Near-RT} \ac{RIC} allow the \acp{MNO} to monitor live network performance metrics via the \ac{E2SM-KPM} \cite{E2KPM} and to change the network parameters and configurations via the \ac{E2SM-RC} \cite{E2RC} and \ac{E2SM-CCC} \cite{E2CCC} to optimize the network performance and \ac{UE} \ac{QoS}. Such standardized methods for monitoring and controlling the network have attracted many interests from the academia and industry and resulted in several optimization directions such as energy efficiency \cite{Lance_EE_ORAN_survey_2024}, traffic steering \cite{lacava2023programmable} and network slicing \cite{yeh2023deep}. In this paper, we address the energy efficiency optimization assisted by a powerful industrial \ac{RAN} emulation and \ac{RIC} testing tool, the TeraVM \ac{AI} \ac{RSG} provided by VIAVI \cite{viavi2025teravmAiRSG}. This tool creates a \ac{DT} of the \ac{RAN} that simulates system level behaviour of the network with scalable and flexible deployment options such as a large number of \acp{UE} and cells, 3GPP standardized propagation models, \ac{UE} mobility and traffic profiles, cell \ac{RF} and energy models, \ac{MAC} scheduler algorithms, etc. The \ac{RAN} nodes can be controlled during live simulations via the REST \ac{API} or E2 messages with actuations such as switching cells on and off and issuing \ac{HO} commands. The \ac{RSG} also offers exposing the network \ac{KPM} reports and \ac{RC} commands to external IPs via the E2 interface which makes it the ideal tool for testing \ac{RIC} and xApp development. Later in this paper we will explain in detail how we configure this tool to generate a \ac{DT} for a large-scale network and the amount of energy saving achieved with our \ac{AI}/\ac{ML} powered xApp. The contributions of this work are as follows:
\begin{itemize}
    \item We propose a novel hybrid Energy Saving xApp that integrates heuristic rules and unsupervised machine learning for intelligent O-RU sleep control within Open RAN.
    \item We leverage \ac{DT} via the VIAVI TeraVM AI-RSG to emulate large-scale Open RAN networks with realistic user mobility and channel propagation conditions.
    \item We design lightweight clustering-based mechanisms to identify and activate suitable sleeping cells to meet dynamic user demands, while switching off underutilised RUs with minimal computational overhead.
    \item We validate our approach in a dense urban-like environment with 246 UEs and 51 cells, demonstrating up to 13\% energy savings without compromising user QoS.
\end{itemize}


\section{Architectural Design}
\label{sec:Architectural Design}
\begin{figure}[!t]
	\centering
	\includegraphics[clip, trim=0.0cm 0.cm 0.0cm 0cm, width=0.6\columnwidth]{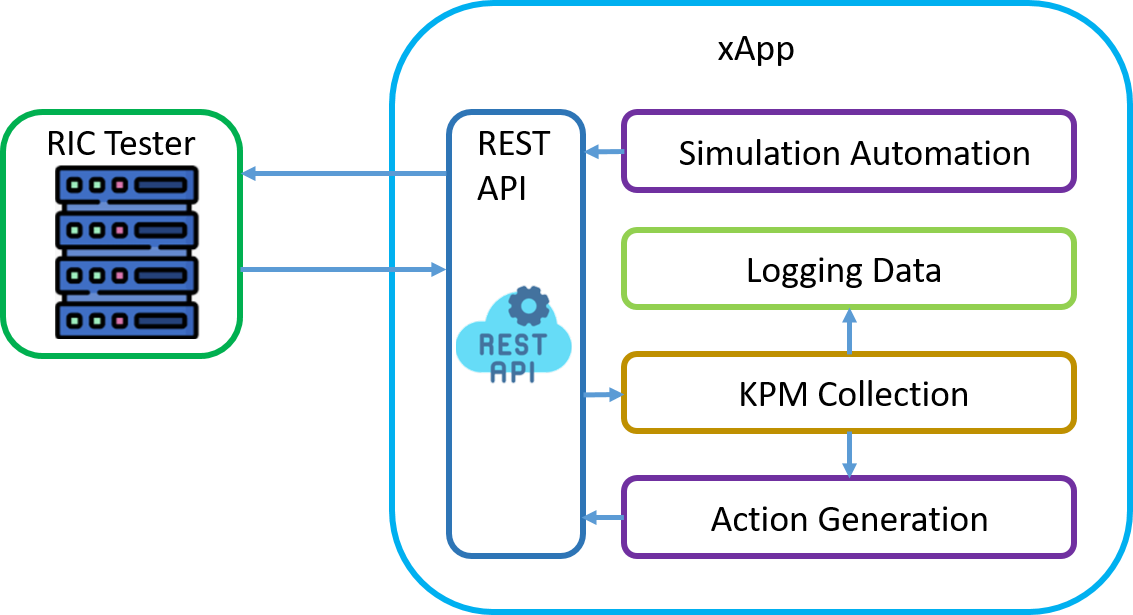}
	\caption{High-level architecture and submodules}
	\label{fig:xapp_archi}
    \vspace{-5mm}
\end{figure}

Fig. \ref{fig:xapp_archi} shows the high-level architectural design of the system, the interactions between the proposed \ac{ES}-xApp and the \ac{AI} \ac{RSG} (RIC Tester) as well as the submodules of the xApp. The functionalities of the submodules are described below.

\begin{itemize}
    \item REST \ac{API}: this submodule manages interactions between the xApp and the RIC tester via HTTP requests sent to specific endpoints and corresponding responses. Its detailed functionalities are described in submodules below.
    \item Simulation Automation: this submodule has two interactions with the RIC tester including starting the network simulation using HTTP POST (with a payload) with a specific network configuration (json format) and stopping the network simulation using HTTP DELETE. This submodule allows starting/stopping multiple simulations sequentially with predefined configuration files, which is particularly useful for testing xApps with the simulations of different seed values (simulations with the same seed will have exactly the same behaviour). 
    \item Logging Data: this submodule creates log files (.csv) for each simulation. Four separate log files are created to record \textit{Cell Reports}, \textit{UE Serving Cell Reports}, \textit{UE Neighbour Cell Reports} and \textit{Aggregated Performance Results} with timestamps. The first three are identical to the KPM data obtained from the KPM Collection submodule and the last one includes processed performance results such as total network power consumption, total/average UE throughput, number of UEs experiencing throughput outage, number of MACRO/MICRO cells under heavy \ac{PRB} utilisation, number of MICRO cells with no UE connected to and number of UEs not requesting any throughput. 
    \item KPM Collection: this submodule uses HTTP GET to query the influxDB of the RIC tester for live \textit{Cell Reports}, \textit{UE Serving Cell Reports} and \textit{UE Neighbour Cell Reports} once or multiple times periodically depending on the requirements of the xApp. The urls for the HTTP GET are carefully tuned so the returned reports contain the latest KPMs for each UE/cell. We also implemented data integrity checking (as the returned KPMs may contain “NaN” in some fields) and  data duplication checking. The collected KPMs are used to support the Logging Data submodule and the Action Generation submodule.
    \item Action Generation: this submodule contains the underlying algorithms for generating commands for changing the cell status (on/off). Depending on the specific xApp, different subsets of the collected KPMs are used as inputs to the heuristic and \ac{ML} algorithms for generating the associated cell on/off commands. HTTP POST will then be used (with the target cell name and action as payload) to execute the actions within the RIC tester. With the configured network scenario, turning on a cell is made effective almost immediately but turning off a cell has a 10 to 20 second delay. 
\end{itemize}

Several options are available for the underlying algorithms for the Action Generation submodule. They have pros and cons in terms of complexity, reliability and comprehensiveness. An overview of these options is provided below 

\begin{itemize}

\item {ML-based xApp:} leverages data-driven intelligence to optimise cell activation and deactivation and can be broadly categorised into three types: {Supervised Learning}, {Unsupervised Learning} and {Reinforcement Learning (RL)}. In {Supervised ML}, models are trained on labelled datasets (ground truth) to predict network metrics such as per-cell or per-area throughput. These predictions are then used to decide whether to turn specific cells on or off. However, acquiring reliable ground truth in dynamic network environments is difficult due to constantly changing conditions like user mobility, interference and load distribution. \textbf{Unsupervised ML}, which is the approach adopted in this work, eliminates the need for labelled data by identifying hidden patterns in \ac{KPM} data collected from the RIC tester. It can cluster UEs exhibiting similar behaviour to determine which MICRO cells are optimal to activate or deactivate. This approach is preferred due to its lower complexity, flexible data requirements and suitability for real-time applications. In contrast, \textbf{RL}-based xApps rely on interacting with the network environment by performing cell on/off actions and learning from the resulting changes in observed \acp{KPM}. However, in large-scale networks with numerous cell combinations, RL training requires a substantial amount of interaction with the environment to converge to an effective policy, which can be computationally intensive and time-consuming.

\item Heuristic xApp: a non-ML-based xApp which follows pre-defined logic to turn the MICRO cells on/off. Compared with ML-based xApp, the heuristic xApp is very simple to operate and does not require training and is not computational intensive. Depending on how the internal logic is designed, it may not provide the optimal performance but should offer good stability. In a practical environment the heuristic xApp could be used as a fail-safe solution given its predictable behaviour when the decisions made by ML-based xApps are not applicable. Later in this paper we will implement a heuristic xApp for benchmarking the performance.    

\item Hybrid Heuristic-ML-based xApp: this category combines the strengths of both heuristic and ML approaches. In such xApps, certain decisions are made using ML models, such as detecting patterns, clustering UEs, or predicting network load while final actions (e.g., selecting specific cells to switch off) may follow rule-based heuristics that ensure system safety and compliance with constraints. This hybrid approach provides a balanced trade-off between adaptability and stability, enabling intelligent yet controlled decision making. 
Hybrid xApps are particularly suited to real-world deployments where lightweight, explainability and bounded behaviour are essential. Our Hybrid ES-xApp implementation will be discussed later in the paper.

\end{itemize}

\section{Digital twin network model}
\label{sec:Digital_twin}
\begin{figure*}[htbp]
	\centering
	\includegraphics[width=0.82\linewidth]{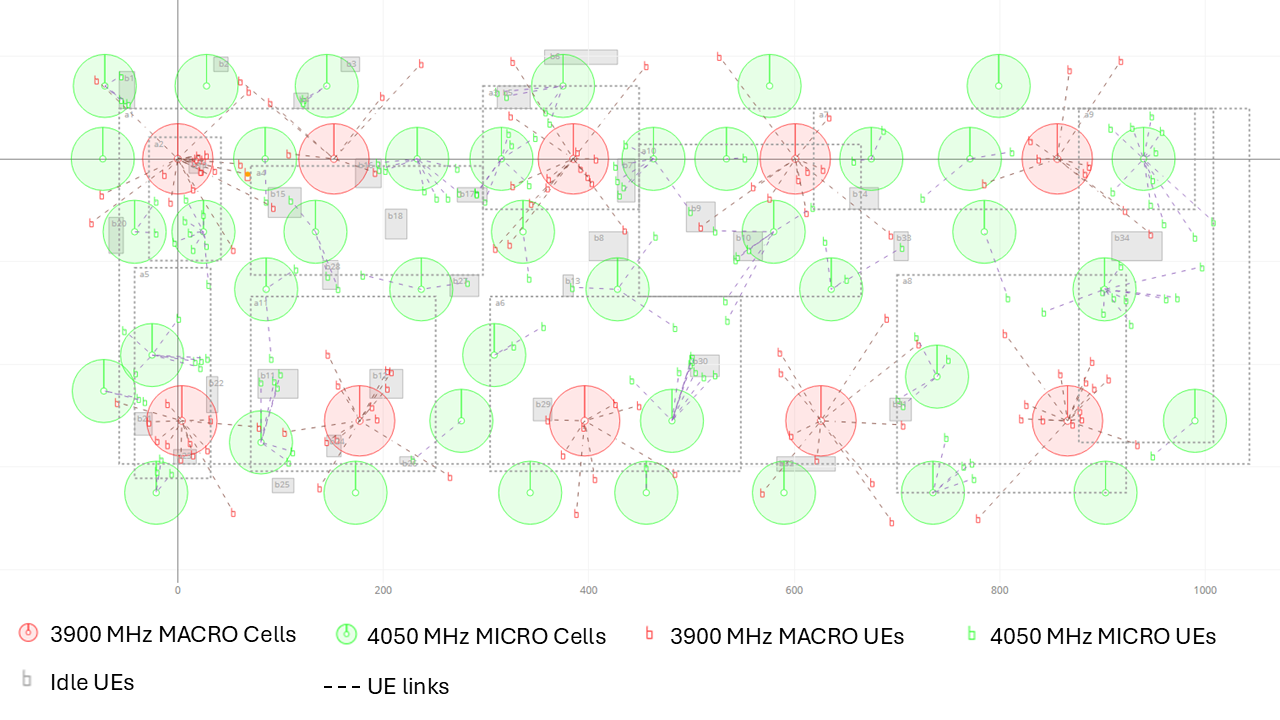}
     \vspace{-3mm}
	\caption{Emulated network scenario}
	\label{fig:RSG_network}
 \vspace{-5mm}
\end{figure*}

In this paper, we investigate a significantly larger scale network scenario compared to our previous works \cite{Qiao_EE_camad_2024, Lance_wincom_EE_xApp_2024} to evaluate the performance of the proposed \ac{ES}-xApp within a more realistic \ac{DT}-based environment provided by the TeraVM AI RSG \cite{viavi2025teravmAiRSG}. Fig \ref{fig:RSG_network} shows the emulated network scenario generated within the AI RSG.

Two types of cells are included in the emulated network (in a $1.2\times0.6$ km area), the MACRO cells (red circles) which provide large area coverage and the MICRO cells (green circles) which provide capacity boosting. The MACRO cells are always on for the connectivity of mobile UEs and the MICRO cells can be switched on/off by an xApp. Table \ref{cell_configurations} lists the configured parameters for the MACRO and MICRO cells. Note that we have configured the MICRO cells to be switched on almost immediately but with a delay for switching off. Once a switching off commands is issued to a specific cell, the \ac{RF} output power of the cell is gradually reduced so that the UEs it serves (if there are any) can be \ac{HO}ed to neighbouring cells.

In Fig. \ref{fig:RSG_network} the UEs connected to each type of cells (at the moment of the snapshot) are marked with the cells' associated colour. A total of 246 UEs are configured with 4 types of mobility models, including static in-building UEs (in grey boxes), pedestrian UEs (2 m/s), UEs in slow cars (10 m/s) and UEs in fast cars (15/m). Every type of UEs have a specific mobility model and traffic profile which are highlighted in Table \ref{table:UEs_configurations}. All mobile UEs are outdoor with an average heigh of 1.5 m above the ground and the static UEs are in buildings (grey boxes) with various heights between 20 and 50 meters.


\begin{table}[t!]
\centering
\vspace{\baselineskip}  
\caption{MACRO/MICRO Cell Configurations}
\resizebox{\linewidth}{!}{%
\begin{tabular}{|>{\raggedright\arraybackslash}m{5.2cm}|>{\raggedright\arraybackslash}m{4.5cm}|}
\hline
\textbf{Configuration Item} & \textbf{Value (MACRO / MICRO)} \\ \hline
Number of cells & 10 / 41 \\ \hline
Center frequency & 3900 / 4050 MHz \\ \hline
Channel model & UMa / UMi \\ \hline
Bandwidth & 100 MHz \\ \hline
RF output power & 45 / 32 dBm \\ \hline
Antenna height & 20 / 10 m \\ \hline
Antenna tilt & 10\degree{} / 5\degree{} \\ \hline
Antenna type & Isotropic \\ \hline
Max. power consumption & 379 / 172 W \\ \hline
Sleep state power consumption & NA / 8 W \\ \hline
Cell shutdown delay & NA / 10 s \\ \hline
Power reduction rate (shutdown) & NA / 3 dB/s \\ \hline
\end{tabular}%
}
\label{cell_configurations}
\vspace{-2mm}
\end{table}

\begin{table}[t]
\caption{UE configurations}
\resizebox{\columnwidth}{!}{%
\begin{tabular}{|l|l|l|l|l|l|}
\hline
\textbf{UE type} &
  \textbf{Number} &
  \textbf{\begin{tabular}[c]{@{}l@{}}Speed \\ (m/s)\end{tabular}} &
  \textbf{\begin{tabular}[c]{@{}l@{}}Target throughput\\ (Mbps)\end{tabular}} &
  \textbf{\begin{tabular}[c]{@{}l@{}}Average time \\ between calls (s)\end{tabular}} &
  \textbf{\begin{tabular}[c]{@{}l@{}}Average call \\ duration (s)\end{tabular}} \\ \hline
Pedestrian & 64 & 2  & 20 & 1000 & 30 \\ \hline
Indoor     & 50 & -  & 50 & 600  & 30 \\ \hline
Fast car   & 75 & 15 & 30 & 100  & 30 \\ \hline
Slow car   & 57 & 10 & 23 & 600  & 30 \\ \hline
\end{tabular}%
}
\label{table:UEs_configurations}
\vspace{-5mm}
\end{table}

\section{Problem Formulation}
The objective of our energy-saving strategy is to minimise overall energy consumption by optimally managing the ON–OFF states of RUs, selectively deactivating underutilised MICRO RUs, while preserving network performance and maintaining \ac{QoS}. The total number of MICRO \acp{RU} that are turned off is denoted by \(\mathcal{Z}\).
The network consists of two classes of \acp{RU}: MACRO RUs, which are always active and MICRO \acp{RU}, which are dynamically controlled by the xApp. We define \(\mathcal{M}_\text{macro}\) and \(\mathcal{M}_\text{micro}\) denote the sets of MACRO and MICRO RUs respectively, and \(\mathcal{M} = \mathcal{M}_\text{macro} \cup \mathcal{M}_\text{micro}\) be the set of all RUs. Similarly, let \(\mathcal{U}\) denote the set of User Equipments (UEs) in the network.

Let \(\alpha_{k,m} \in \{0,1\}\) be a binary indicator variable representing whether UE \(k \in \mathcal{U}\) is connected to RU \(m \in \mathcal{M}\), and \(s_m \in \{0,1\}\) be the operational status of RU \(m\), where \(s_m = 1\) indicates that the RU is active.
We define the optimisation problem as:

\begin{subequations} \label{eq:optimization_problem}
\vspace{-3mm}
\begin{align}
\max_{s_m} \quad & \mathcal{Z} = \sum_{m \in \mathcal{M}_\text{micro}} (1 - s_m) 
    && \label{eq:objective} \\[1ex]
\text{s.t.} \quad 
& \sum_{m \in \mathcal{M}} \alpha_{k,m} = 1, 
    && \forall k \in \mathcal{U} \label{eq:association} \\
& \sum_{m \in \mathcal{M}} \alpha_{k,m} R_{k,m} \geq R_\text{min}, 
    && \forall k \in \mathcal{U} \label{eq:rss} \\
& \sum_{k \in \mathcal{U}} \alpha_{k,m} \leq C_\text{max}, 
    && \forall m \in \mathcal{M} \label{eq:capacity} \\
& \alpha_{k,m} \leq s_m, 
    && \forall k \in \mathcal{U}, \; \forall m \in \mathcal{M}_\text{micro} \label{eq:status_constraint} \\
& s_m = 1, 
    && \forall m \in \mathcal{M}_\text{macro} \label{eq:macro_constraint}
\end{align}
\end{subequations}

where $R_{min}$ denotes the minimal acceptabale signal power and \( C_{\max} \) represents the maximum number of UEs that each \ac{RU} can serve concurrently.
Constraints \eqref{eq:association} and \eqref{eq:rss} ensure that each UE is assosiated with a single RU and receives a minimal required signal quality. Constraint \eqref{eq:capacity} enforces RU capacity limits, while \eqref{eq:status_constraint} ensures that UEs are only associated with active MICRO RUs when needed. MACRO RUs are always ON as enforced by \eqref{eq:macro_constraint}.

Given the NP-hardness of this mixed-integer optimisation, we propose a hybrid solution integrating lightweight unsupervised learning (for cell activation) and heuristics (for deactivation) to achieve near-optimal results in large-scale emulated Open RAN environments.

\section{\ac{ES}-xApp Design}
The proposed lightweight \ac{ES}-xApp combines a heuristic component and an \ac{ML} component to tackle the complicated network scenario. The identification of underutilised cells is handled through a heuristic approach (switching off), while the ML component detects capacity-demanding areas and activates sleeping cells when needed. Algorithm \ref{alg:rule_based_xapp} summarises the proposed ES-xApp.

\subsection{ML Component}
The ML component of the proposed ES-xApp employs an unsupervised learning strategy to assist in cell activation when network congestion is detected. Specifically, we apply the K-Means clustering algorithm to spatially group sleeping cells and active UEs, enabling identification of the most suitable cell to switch on.
The K-Means algorithm partitions \acp{UE} and sleeping cells based on their coordinates into $K$ spatial clusters. The clustering aims to minimise the total within-cluster variance, formulated as:
\begin{equation} \label{eq:kmeans}
\underset{\{\mathcal{C}_k\}_{k=1}^K}{\text{min}} \sum_{k=1}^{K} \sum_{\mathbf{x}_i \in \mathcal{C}_k} \| \mathbf{x}_i - \boldsymbol{\mu}_k \|^2,
\end{equation}
where $\mathcal{C}_k$ is the $k$-th cluster, $\mathbf{x}_i$ is a 2D position vector (of either a UE or a sleeping cell) and ${\mu}_k$ is the centroid of cluster $k$. The key steps in the ML logic are as follows:
\begin{itemize}
    \item Clustering for cell activation: identifies the most suitable cell to activate when a nearby cell is overloaded. 
    \item Weighted distance calculation: for each cluster, the algorithm calculates the weighted distance between UEs and the sleeping cell in the cluster with the weighting factor prioritizing UEs with higher throughput demands.
        \begin{equation} \label{eq:weighted_dist}
    D_k = \sum_{i=1}^{N_k} w_i \cdot \| \mathbf{x}_i - \mathbf{c}_k \|,
    \end{equation}
    where $N_k$ is the number of \acp{UE} in cluster $k$, $\mathbf{x}_i$ is the position of UE $i$, $\mathbf{c}_k$ is the position of the sleeping cell in cluster $k$, and $w_i$ is the throughput demand of UE $i$.

    \item Cell selection: the sleeping cell  with the lowest weighted distance $D_k$ is selected for activation. 
    \begin{equation} \label{eq:best_cell}
            k^* = \arg\min_k D_k.
    \end{equation}
    This  ensures the activated cell serves the most demanding UEs in terms of throughput, located closest to it. 
\end{itemize}

\subsection{Heuristic Component}
The heuristic component relies on predefined policies to make switching-off decisions. These rules are based on the reported KPMs collected from the network. The key steps in the heuristic logic are:
\begin{itemize}
    \item Switch off idle cells: for every round of KPM collection, the cells with no connected UEs are identified as idle cells $C_{\text{idle}}$ as:
    \begin{equation}
        C_{\text{idle}} = \left\{ c \in C : \text{ConnMean}(c) = 0 \right\},
    \end{equation}
    where $\text{ConnMean}(c)$ is the average number of connected \acp{UE} in cell $c$ over a monitoring period. These cells are placed into sleep mode unless they were recently switched on and are still within a protection timer $T_{\text{on}}$.

\item Threshold-based decision for low PRB utilisation: a MICRO cell $c$ with low downlink \ac{PRB} usage is eligible for switch-off if $\text{PRB}_{\text{DL}}(c) < \rho$, where $\rho$ is the PRB utilisation threshold (50\% in our case).
    In addition, for each UE $u$ served by $c$, the following two conditions must be met for by least a single neighbouring cell $c'$:
    \begin{equation}
    \text{PRB}_{\text{DL}}(c') < \rho \quad \text{and} \quad \text{RSRP}(u, c') > R_{\min},
    \end{equation}
    where $\text{RSRP}(u, c')$ is the reference signal received power of UE $u$ from neighbouring cell $c'$, and $R_{\min}$ is the RSRP threshold (-110 dBm in our case). If all the conditions are met and all UEs served by $c$ can be handed over to neighbouring cells, then $c$ is placed into sleep mode.

    
\end{itemize}

\begin{algorithm}[!t]
\DontPrintSemicolon
  \KwInput{Real-time cell metrics $C$, UE reports $U$, neighbour reports $N$}
  \KwOutput{Cell activation/deactivation commands}

  Initialise list of active cells $C_\text{on}$ and sleeping cells $C_\text{sleep}$ \\
  Define thresholds: PRB utilisation $\rho$, RSRP minimum $r_{min}$, distance max $d_{max}$ \\
  
  \While{within xApp runtime}{
    
    Fetch latest $C$, $U$, and $N$ from the simulator \\
    Filter out invalid or incomplete entries in $C$, $U$, and $N$ \\

    \tcp{--- Switch-On Logic ---}
    Identify overloaded cells $C_{\text{over}}$ where PRB utilisation = 100 \\
    \ForEach{$c \in C_{\text{over}}$}{
        Identify UEs connected to $c$, denoted $U_c$ \\
        Identify $C_{\text{sleep}}^{near}$ within $d_{max}$ of $c$ \\
        \If{$C_{\text{sleep}}^{near} \neq \emptyset$}{
            Construct data matrix of coordinates from $U_c$ and $C_{\text{sleep}}^{near}$ \\
            Apply KMeans clustering with $k = |C_{\text{sleep}}^{near}|$ \\
            \ForEach{cluster}{
                Compute weighted distance between sleeping cell and UEs based on UE throughput \\
            }
            Select sleeping cell with minimum weighted distance and switch it on \\
            Update activation timestamp for this cell
        }
    }

    \tcp{--- Switch-Off Logic ---}
    Identify cells in $C$ with no active UEs and not in energy-saving mode \\
    \If{such cells exist}{
        Select one such cell randomly and switch it off \\
    }
    \Else{
        Identify lightly loaded cells: PRB usage $< \rho$ and not in energy-saving mode \\
        \ForEach{cell $c$ in lightly loaded set}{
            Identify UEs served by $c$ \\
            For each UE, find neighbour cells with PRB usage $< \rho$ and RSRP $\geq r_{min}$ \\
            \If{handover is feasible for all UEs}{
                Switch off cell $c$ \\
                \textbf{break}
            }
        }
    }

    Sleep for a short duration before next iteration
  }
  \caption{Cell Power Management Algorithm}
  \label{alg:rule_based_xapp}
\end{algorithm}

\section{Results}
\label{results}
To evaluate the performance of the proposed energy-saving hybrid xApp, we conducted extensive experiments using the VIAVI TeraVM AI RSG to emulate a realistic dense urban Open RAN scenario utilising the VIAVI TeraVM AI RSG. In addition, we compare the proposed ES-xApp with a heuristic ES-xApp. The heuristic xApp switches off any MICRO cells without \acp{UE} attached to it. As for the switch on, it checks if a MACRO cell is heavily utilised (> 90\% PRB usage), a random MICRO cell (within it's vicinity) is turned back on. The emulation includes a mixture of MACRO and MICRO O-RUs with different types of UEs (i.e., indoor UEs, pedestrians, slow moving vehicles and fast moving vehicles), Table \ref{table:UEs_configurations} highlights the \acp{UE} characteristics. The performance of the proposed approach was benchmarked against a heuristic energy saving xApp that switches MICRO cells into sleep mode whenever they don't have \acp{UE} attched to them. While the switch on mechaism is when a MACRO cell have full \ac{PRB} utilisation, it switches on a MICRO cell within it's vicinity.

The experiments were conducted over a 2-hour duration, Tabled \ref{tab:power_reduction} shows the averaged power consumption and averaged downlink throughput by the digital twin of the O-RAN network.
With All ON: Serving as the baseline, this scenario consumes the highest power at 4.87 KW with a downlink throughput of 
2.47 Gbps. This represents the case where all MICRO cells are continuously active and no power-saving measures are employed.
The heuristic-based xApp reduces power consumption moderately to 4.53 kW, achieving approximately 6.98$\%$ savings compared to the baseline. However, this comes with a noticeable 3.32$\%$ decrease in throughput (2.39 Gbps), indicating the limitations of heuristic only approach in balancing between energy efficiency and network performance. The proposed hybrid heuristic-ML xApp further improves energy savings, reducing power consumption to 4.32 kW, which corresponds to 13.27\% savings compared to the baseline. Notably, this is achieved while maintaining a near baseline throughput of 2.46 Gbps, resulting in only 0.4\% degradation.
The obtained results highlight the improvement in energy saving with a minimal impact on the \acp{UE} quality of service in large-scale O-RAN deployments.

\begin{figure*}[htbp]
	\centering
	\includegraphics[clip, trim=0.0cm 7.cm 0.0cm 6cm, width=1.92\columnwidth]{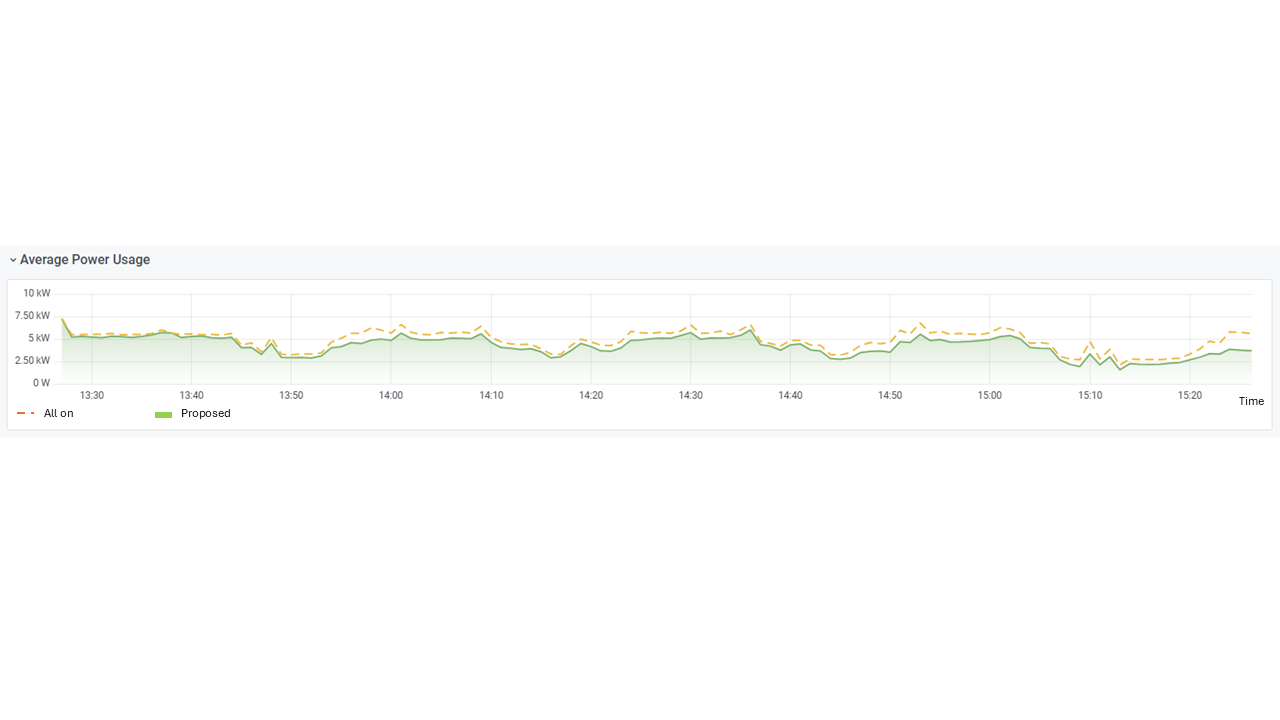}
        \vspace{-3mm}
    \caption{Average power usage comparison. The simulation consistently shows lower energy usage compared to the baseline, demonstrating energy-saving effectiveness.}
    \label{fig:power_usage}
            \vspace{-1mm}
\end{figure*}

\begin{figure*}[htbp]
	\centering
	\includegraphics[clip, trim=0.0cm 7.cm 0.0cm 6cm, width=1.92\columnwidth]{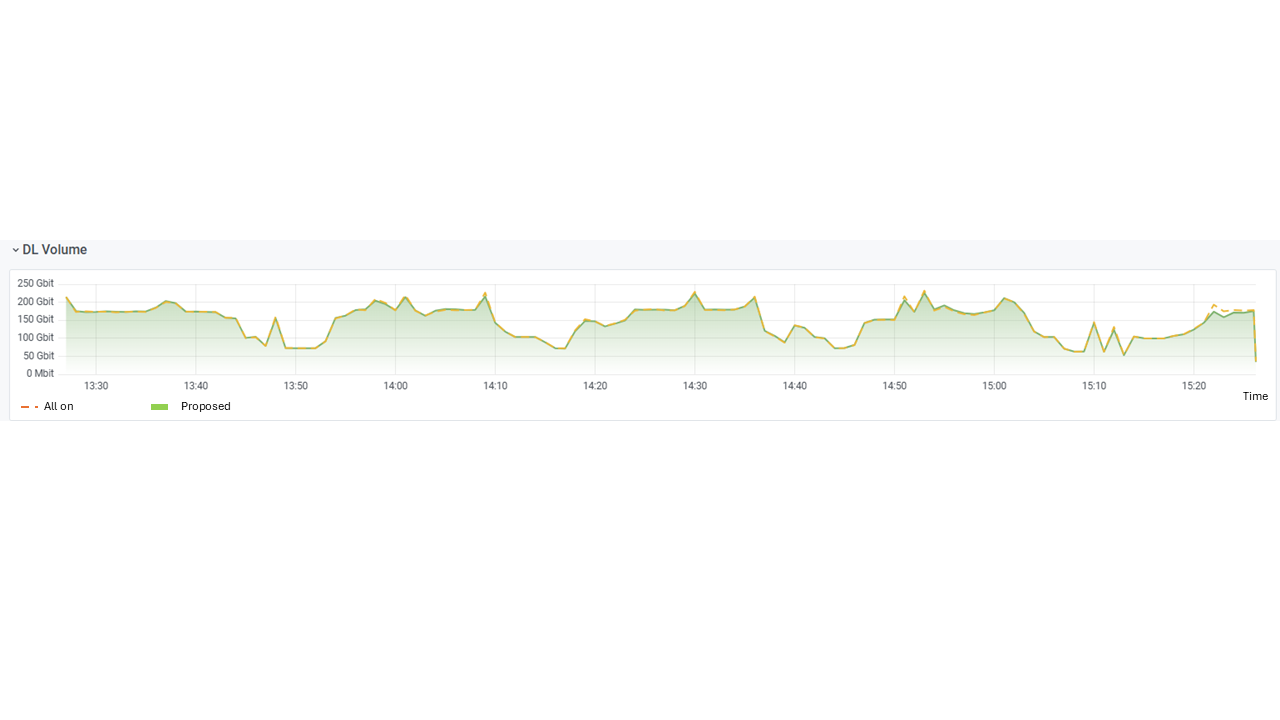}
            \vspace{-3mm}
    \caption{Downlink (DL) volume comparison. The simulation maintains comparable throughput to the baseline, indicating performance is not compromised.}
    \label{fig:dl_volume}
                \vspace{-1mm}
\end{figure*}

Next, Fig. \ref{fig:power_usage} shows the power usage by the proposed simulation compared to the baseline scenario across the monitored simulation timeframe. The dashed baseline line represents the scenario in which all cells remain active continuously, reflecting higher power consumption. In contrast, the solid simulation line represents the proposed hybrid xApp consistently stays below the baseline, confirming that the proposed approach effectively reduces power consumption. This sustained energy-saving performance highlights the capability of the hybrid approach to manage the network’s energy resources dynamically. While Fig. \ref{fig:dl_volume} shows the DL throughput for the baseline and proposed method. Notably, despite the power reduction achieved by the proposed simulation, the DL throughput remains closely aligned with the baseline throughout the entire simulation period. This indicates that the hybrid xApp maintains robust network performance, ensuring that the energy-saving measures do not adversely impact the user experience. The small throughput variations observed are within an acceptable range, suggesting minimal QoS trade-offs. In addition, small dips in throughput are observed; however, these correlate with periods of reduced traffic demand, during which the algorithm switches off unnecessary MICRO cells, resulting in proportional energy savings. Overall, these results reinforce the effectiveness of the proposed hybrid xApp in simultaneously achieving substantial energy savings and maintaining high-quality network performance.


\begin{table}[]
\caption{Power consumption and throughput comparison across methods.}
\label{tab:power_reduction}
\begin{tabular}{|l|l|l|l|l|}
\hline
\textbf{Method}                                             & \textbf{\begin{tabular}[c]{@{}l@{}}Power \\ (kW)\end{tabular}} & \textbf{\begin{tabular}[c]{@{}l@{}}Reduction \\ (\%)\end{tabular}} & \textbf{\begin{tabular}[c]{@{}l@{}}DL Throughput \\ (Gbps)\end{tabular}} & \textbf{\begin{tabular}[c]{@{}l@{}}Reduction \\ (\%)\end{tabular}} \\ \hline
\begin{tabular}[c]{@{}l@{}}All ON\\ (baseline)\end{tabular} & 4.87                                                           & --                                                                 & 2.47                                                                     & --                                                                 \\ \hline
Heuristic                                                   & 4.53                                                           & 6.98                                                               & 2.39                                                                     & $-3.32$                                                            \\ \hline
Proposed                                                    & 4.32                                                           & 13.27                                                              & 2.46                                                                     & $-0.4$                                                             \\ \hline
\end{tabular}
\vspace{-5mm}
\end{table}

\section{Conclusions}
In this work, we have introduced a novel hybrid xApp that integrates heuristic methods with unsupervised machine learning, supported by digital twin technology for intelligent energy management in large-scale Open RAN networks. The proposed xApp dynamically controls the sleep modes of Open Radio Units, achieving significant energy savings while preserving user Quality of Service. Our evaluations using the TeraVM AI RAN Scenario Generator confirm that this method can deliver approximately 13$\%$ energy reduction in a realistic emulated environment. These results highlight the feasibility and effectiveness of lightweight AI-driven hybrid approaches to address the critical challenge of energy efficiency in next-generation wireless networks.





\input{acro}

\footnotesize
\bibliographystyle{IEEEtran}
\bibliography{References}

\end{document}

%% file: acro.tex
\begin{acronym} 
\acro{3GPP}{3rd Generation Partnership Project}
\acro{5G}{Fifth Generation}
\acro{AI}{Artificial Intelligence}
\acro{ANN}{Artificial Neural Network}
\acro{API}{Application Programming Interface}
\acro{BBU}{Base Band Unit}
\acro{BER}{Bit Error Rate}
\acro{BS}{Base Station}
\acro{BW}{bandwidth}
\acro{C-RAN}{Cloud Radio Access Networks}
\acro{CDMA}{Code Division Multiple Access}
\acro{CNN}{Conventional Neural Network}
\acro{CoMP}{Coordinated Multipoint}
\acro{COTS}{Commercial off-the-shelf}
\acro{CP}{Control Plane}
\acro{CR}{Cognitive Radio}
\acro{CU}{Central Unit}
\acro{DAC}{Digital-to-Analog Converter}
\acro{D2D}{Device-to-Device}
\acro{DAS}{Distributed Antenna Systems}
\acro{DC}{Duty Cycle}
\acro{DDQN}{Double Deep Q-Learning Network}
\acro{DL}{Deep Learning}
\acro{DNN}{Deep Neural Network}
\acro{DQN}{Deep Q-Learning Network}
\acro{DRL}{Deep Reinforcement Learning}
\acro{D-RAN}{Distributed RAN}
\acro{DSA}{Dynamic Spectrum Access}
\acro{DT}{Digital Twin}
\acro{DU}{Distributed Unit}
\acro{EE}{Energy Efficiency}
\acro{eMMB}{enhanced Mobile Broadband}
\acro{E2SM-KPM}{E2 Service Model - Key Performance Measurements}
\acro{E2SM-RC}{E2 Service Model - RAN Control}
\acro{E2SM-CCC}{E2 Service Model - Cell Configuration Control}
\acro{ES}{Energy Saving}
\acro{FDD}{Frequency Division Duplex}
\acro{FEC}{Forward Error Correction}
\acro{FFT}{Fast Fourier Transform}
\acro{FFR}{Fractional Frequency Reuse}
\acro{FPGAs}{Field  Programmable  Gate  Arrays}
\acro{FSPL}{Free Space Path Loss}
\acro{GA}{Genetic Algorithms}
\acro{GNN}{Graph Neural Network}
\acro{GUI}{Graphical User Interface}
\acro{HARQ}{Hybrid-Automatic Repeat Request}
\acro{HDD}{High Density Deployment}
\acro{HetNet}{Heterogeneous Network}
\acro{HO}{Handover}
\acro{HTTP}{Hypertext Transfer Protocol}
\acro{IC}{Immersive Communication}
\acro{ICA}{Independent Component Analysis}
\acro{ILP}{Integer Linear Programming}
\acro{IoT}{Internet of Things}
\acro{KNN}{k-Nearest Neighbors}
\acro{KPM}{Key Performance Measurements}
\acro{KPI}{Key Performance Indicators}
\acro{LAN}{Local Area Network}
\acro{LOS}{Line of Sight}
\acro{LTE}{Long Term Evolution}
\acro{LTE-A}{Long Term Evolution Advanced}
\acro{LSTM}{Long Short-term Memory}
\acro{MAC}{Medium Access Control}
\acro{MDP}{Markov Decision Process}
\acro{ML}{Machine Learning}
\acro{MLP}{Multiple-layer Perceptron}
\acro{mMTC}{massive Machine-Type Communication}
\acro{MIMO}{Multiple-Input Multiple-Output}
\acro{m-MIMO}{massive Multiple-Input Multiple-Output}
\acro{mmWave}{millimeter Wave}
\acro{MNO}{Mobile Operator}
\acro{Near-RT}{Near-Real Time}
\acro{Near-RT RIC}{Near Real Time RIC}
\acro{NFV}{Network Function Virtualization}
\acro{NIB}{Network Information Base}
\acro{NLoS}{Non-Line of Sight}
\acro{NN}{Neural Network}
\acro{NR}{New Radio}
\acro{OFDM}{Orthogonal Frequency Division Multiplexing}
\acro{OFDMA}{Orthogonal Frequency-Division Multiple Access}
\acro{O-RAN}{Open Radio Access Network }
\acro{Open RAN}{Open Radio Access Network}
\acro{OPEX}{Operational Expenditure}
\acro{OSC}{O-RAN Software Community}
\acro{OSA}{Opportunistic Spectrum Access}
\acro{PA}{Power Amplifier}
\acro{PAM}{Pulse Amplitude Modulation}
\acro{PAPR}{Peak-to-Average Power Ratio}
\acro{PCA}{Principal Component Analysis}
\acro{PDCP}{Packet Data Convergence Protocol}
\acro{PDU}{Power Distribution Unit}
\acro{PG}{Policy Gradient}
\acro{PHY}{Physical layer}
\acro{PRB}{Physical resource block}
\acro{PSO}{Particle Swarm Optimization}
\acro{PU}{Primary User}
\acro{QL}{Q-Learning}
\acro{QAM}{Quadrature Amplitude Modulation}
\acro{QoE}{Quality of Experience}
\acro{QoS}{Quality of Service}
\acro{QPSK}{Quadrature Phase Shift Keying}
\acro{RAN}{Radio Access Network}
\acro{RAT}{Radio Access Technology}
\acro{RC}{RAN Control}
\acro{RF}{Radio Frequency}
\acro{RIC}{RAN Intelligent Controller}
\acro{RLC}{Radio Link Control}
\acro{RL}{Reinforcement Learning}
\acro{RMSE}{Root Mean Squared Error}
\acro{RN}{Remote Node}
\acro{RRH}{Remote Radio Head}
\acro{RRM}{Radio Resources Management}
\acro{RRC}{Radio Resource Control}
\acro{RRU}{Remote Radio Unit}
\acro{RSG}{RAN Scenario Generator}
\acro{RSS}{Received Signal Strength}
\acro{RSRP}{Reference Signals Received Power}
\acro{RT}{Real Time}
\acro{RU}{Radio Unit}
\acro{SCA}{Successive Convex Approximation}
\acro{SCBS}{Small Cell Base Station}
\acro{SDL}{Shared Data Layer}
\acro{SDN}{Software Defined Network}
\acro{SDR}{Software Defined Radio}
\acro{SDS}{Software Defined Security}
\acro{SDAP}{Service Data Adaptation Protocol}
\acro{SHAP}{SHapley Additive exPlanations}
\acro{SINR}{Signal-to-Interference-plus-Noise Ratio}
\acro{SLA}{Service Level Agreement}
\acro{SMO}{Service Management and Orchestration}
\acro{SNR}{Signal-to-Noise Ratio}
\acro{SON}{Self-organised Network}
\acro{SU}{Secondary User}
\acro{SU-MIMO}{Single-User MIMO}
\acro{SVM}{Support Vector Machine}
\acro{TD}{Temporal Defence}
\acro{TDD}{Time Division Duplex}
\acro{TD-LTE}{Time Division LTE}
\acro{TDMA}{Time Division Multiple Access}
\acro{TS}{Traffic Steering}
\acro{UDN}{Ultra-Dense Network}
\acro{UE}{User Equipment}
\acro{UMa}{Urban Macrocell path loss}
\acro{UMi}{Urban Microcell path loss}
\acro{URLLC}{ultra-reliable low-latency communication}
\acro{USRP}{Universal Software Radio Platform}
\acro{VFN}{Network Function Virtualization}
\acro{v-RAN}{Virtual RAN}
\acro{XAI}{eXplainable Artificial Intelligent}
\end{acronym}

%% file: submitted_6pages.bbl
\begin{thebibliography}{10}
\providecommand{\url}[1]{#1}
\csname url@samestyle\endcsname
\providecommand{\newblock}{\relax}
\providecommand{\bibinfo}[2]{#2}
\providecommand{\BIBentrySTDinterwordspacing}{\spaceskip=0pt\relax}
\providecommand{\BIBentryALTinterwordstretchfactor}{4}
\providecommand{\BIBentryALTinterwordspacing}{\spaceskip=\fontdimen2\font plus
\BIBentryALTinterwordstretchfactor\fontdimen3\font minus \fontdimen4\font\relax}
\providecommand{\BIBforeignlanguage}[2]{{%
\expandafter\ifx\csname l@#1\endcsname\relax
\typeout{** WARNING: IEEEtran.bst: No hyphenation pattern has been}%
\typeout{** loaded for the language `#1'. Using the pattern for}%
\typeout{** the default language instead.}%
\else
\language=\csname l@#1\endcsname
\fi
#2}}
\providecommand{\BIBdecl}{\relax}
\BIBdecl

\bibitem{feng2012survey}
D.~Feng, C.~Jiang, G.~Lim, L.~J. Cimini, G.~Feng, and G.~Y. Li, ``A survey of energy-efficient wireless communications,'' \emph{IEEE Communications Surveys \& Tutorials}, vol.~15, no.~1, pp. 167--178, 2012.

\bibitem{Imran_EE_survey_2023}
A.~I. Abubakar, O.~Onireti, Y.~Sambo, L.~Zhang, G.~K. Ragesh, and M.~Ali~Imran, ``Energy efficiency of open radio access network: A survey,'' in \emph{2023 IEEE 97th Vehicular Technology Conference (VTC2023-Spring)}, 2023, pp. 1--7.

\bibitem{bonati2021intelligence}
L.~Bonati, S.~D'Oro, M.~Polese, S.~Basagni, and T.~Melodia, ``Intelligence and learning in o-ran for data-driven nextg cellular networks,'' \emph{IEEE Communications Magazine}, vol.~59, no.~10, pp. 21--27, 2021.

\bibitem{ORAN_Architecture_2022}
\BIBentryALTinterwordspacing
{O-RAN Alliance}, ``{O-RAN Architecture Description v03.00},'' O-RAN Technical Specification, 2022, accessed: 2025-03-06. [Online]. Available: \url{https://www.o-ran.org/specifications}
\BIBentrySTDinterwordspacing

\bibitem{RCRWireless_FunctionalSplits_2021}
\BIBentryALTinterwordspacing
{RCR Wireless News}, ``Exploring functional splits in {5G RAN}: Tradeoffs and use cases,'' 2021, accessed: 2025-03-07. [Online]. Available: \url{https://www.rcrwireless.com/20210317/5g/exploring-functional-splits-in-5g-ran-tradeoffs-and-use-cases-reader-forum}
\BIBentrySTDinterwordspacing

\bibitem{polese2023understanding}
M.~Polese, L.~Bonati, S.~D’oro, S.~Basagni, and T.~Melodia, ``Understanding o-ran: Architecture, interfaces, algorithms, security, and research challenges,'' \emph{IEEE Communications Surveys \& Tutorials}, vol.~25, no.~2, pp. 1376--1411, 2023.

\bibitem{RIC}
\BIBentryALTinterwordspacing
{O-RAN Alliance}, ``{O-RAN} near-{RT} {RIC} architecture 4.0,'' ORAN. WG3.RICARCH- R003-v04.00 Technical Specification, Tech. Rep., 2024. [Online]. Available: \url{https://orandownloadsweb.azurewebsites.net/specifications.}
\BIBentrySTDinterwordspacing

\bibitem{O1}
\BIBentryALTinterwordspacing
{O-RAN Alliance}, ``{O-RAN} {O}perations and {M}aintenance {I}nterface 4.0,'' O-RAN.WG1.O1-Interface.0-v04.00 Technical Specification, Tech. Rep., 2021. [Online]. Available: \url{https://orandownloadsweb.azurewebsites.net/specifications.}
\BIBentrySTDinterwordspacing

\bibitem{E2AP}
\BIBentryALTinterwordspacing
{O-RAN Alliance}, ``{O-RAN E2 Application Protocol (E2AP) 7.0},'' O-RAN.WG3.TS.E2AP-R004-v07.00 Technical Specification, Tech. Rep., 2025. [Online]. Available: \url{https://orandownloadsweb.azurewebsites.net/specifications.}
\BIBentrySTDinterwordspacing

\bibitem{E2KPM}
\BIBentryALTinterwordspacing
{O-RAN Alliance}, ``{O-RAN E2 Service Model (E2SM) KPM 6.0},'' O-RAN.WG3.TS.E2SM-KPM-R004-v06.00 Technical Specification, Tech. Rep., 2025. [Online]. Available: \url{https://orandownloadsweb.azurewebsites.net/specifications.}
\BIBentrySTDinterwordspacing

\bibitem{E2RC}
\BIBentryALTinterwordspacing
{O-RAN Alliance}, ``{O-RAN E2 Service Model (E2SM), RAN Control 7.0},'' O-RAN.WG3.TS.E2SM-RC-R004-v07.00 Technical Specification, Tech. Rep., 2025. [Online]. Available: \url{https://orandownloadsweb.azurewebsites.net/specifications}
\BIBentrySTDinterwordspacing

\bibitem{E2CCC}
\BIBentryALTinterwordspacing
{O-RAN Alliance}, ``{E2 Service Model (E2SM) Cell Configuration and Control 5.0},'' O-RAN.WG3.TS.E2SM-CCC-R004-v05.00 Technical Specification, Tech. Rep., 2025. [Online]. Available: \url{https://orandownloadsweb.azurewebsites.net/specifications}
\BIBentrySTDinterwordspacing

\bibitem{Lance_EE_ORAN_survey_2024}
X.~Liang, Q.~Wang, A.~Al-Tahmeesschi, S.~B. Chetty, D.~Grace, and H.~Ahmadi, ``Energy consumption of machine learning enhanced open ran: A comprehensive review,'' \emph{IEEE Access}, vol.~12, pp. 81\,889--81\,910, 2024.

\bibitem{lacava2023programmable}
A.~Lacava, M.~Polese, R.~Sivaraj, R.~Soundrarajan, B.~S. Bhati, T.~Singh, T.~Zugno, F.~Cuomo, and T.~Melodia, ``{Programmable and customized intelligence for traffic steering in 5G networks using Open RAN architectures},'' \emph{IEEE Trans. Mobile Comput.}, vol.~23, no.~4, pp. 2882--2897, 2023.

\bibitem{yeh2023deep}
S.-P. Yeh, S.~Bhattacharya, R.~Sharma, and H.~Moustafa, ``{Deep learning for intelligent and automated network slicing in 5G open RAN (ORAN) deployment},'' \emph{IEEE Open Journal of the Communications Society}, vol.~5, pp. 64--70, 2023.

\bibitem{viavi2025teravmAiRSG}
{VIAVI Solutions}, ``{TeraVM RIC Test},'' Mar. 2025, [Online] Available: \url{https://www.viavisolutions.com/en-uk/products/teravm-ai-rsg}.

\bibitem{Qiao_EE_camad_2024}
\BIBentryALTinterwordspacing
Q.~Wang, S.~Chetty, A.~Al-Tahmeesschi, X.~Liang, Y.~Chu, and H.~Ahmadi, ``{Energy Saving in 6G O-RAN Using DQN-based xApp},'' 2024. [Online]. Available: \url{https://arxiv.org/abs/2409.15098}
\BIBentrySTDinterwordspacing

\bibitem{Lance_wincom_EE_xApp_2024}
X.~Liang, A.~Al-Tahmeesschi, Q.~Wang, S.~Chetty, C.~Sun, and H.~Ahmadi, ``Enhancing energy efficiency in o-ran through intelligent xapps deployment,'' in \emph{2024 11th International Conference on Wireless Networks and Mobile Communications (WINCOM)}, 2024, pp. 1--6.

\end{thebibliography}
